\documentstyle[aps,preprint,eqsecnum,floats,epsf]{revtex}
\tighten
\draft
\begin{document}

\preprint{}
\title{Partition Functions in Statistical Mechanics, Symmetric
Functions, and Group Representations}
\author{A.B. Balantekin\thanks{Electronic address: {\tt
baha@nucth.physics.wisc.edu}}}
\address{Department of Physics, University of Wisconsin\\
         Madison, Wisconsin 53706 USA\thanks{Permanent Address},\\
and\\
Max-Planck-Institut f\"ur Kernphysik,
Postfach 103980, D-69029 Heidelberg, Germany}

\maketitle

\begin{abstract}
Partition functions for non-interacting particles are known to be
symmetric functions. It is shown that powerful group-theoretical
techniques can be used not only to derive these relationships, but
also to significantly simplify calculation of the partition functions
for particles that carry internal quantum numbers. The partition 
function is shown to be a sum of one or more group characters. The
utility of character expansions in calculating the partition functions
is explored. Several examples are given to illustrate these
techniques. 
\end{abstract}


\newpage

\vglue1cm

\section{Introduction}

Although the relationship between partition functions of
non-interacting quantum systems of bosons or fermions and symmetric
functions commonly encountered in the group representation theory has
been 
known for some time (e.g. see Ref. \cite{Balantekin:1991te}) it was
recently highlighted \cite{schmidt} as part of attempts to gain a
deeper understanding of the foundations of the field. In particular
the $N$-particle partition function $Z_N$ for a non-interacting gas is
a complete homogeneous symmetric function of the exponentials of the
single-particle energies for bosons and an elementary symmetric
function of the same for fermions (Definitions and a list of some
properties of these functions are given in the Appendix A). Such
relationships can be useful, for example, in one-dimensional fermionic
systems since in one space dimension interacting fermions can be
considered as non-interacting bosons \cite{luttinger1963}. Spectral
equivalence of bosons and fermions in one-dimensional harmonic
potentials \cite{schmidt1999,crescimanno} and previously noted
recursion relations connecting partition functions with different
numbers of particles \cite{borrmann1993,brosens1997} can be shown to
be consequences of this identification.

The aim of this paper is first to provide a simple group-theoretical
proof  for $Z_N$ being a symmetric function and then to expand this
result 
to some cases of interacting particles and to systems with both bosons
and fermions (supersymmetric systems). In the next section we  first
provide a direct combinatorial proof, then show how that proof follows
from group representation theory. In that section we also show that
for mixed systems of bosons and fermions partition functions become
graded symmetric functions which are the characters of
superalgebras. In Section III, we show that the identification of
$N$-particle partition functions with symmetric functions coupled with
character expansion techniques significantly simplify calculation of
partition functions for particles that carry internal quantum
numbers. In Section IV we show that it is possible to utilize these
techniques to calculate the partition functions for some simple
interacting systems. Finally Section V includes a brief discussion of
our results. 

\section{The relationship between partition function and group
characters}
 
One can easily write down the $N$-particle partition function $Z_N$
for non-interacting particles 
\begin{equation}
\label{1}
Z_N = \sum_{n_1}\sum_{n_2} \cdots \left[ \prod_i x_i^{n_i} \right]
\delta (N - \sum_j n_j), 
\end{equation}
as was written e.g. in the treatment of pion multiplicity
distributions in heavy-ion collisions \cite{Kauffmann:1978vw}. 
Here $x_i = \exp{(-\beta \epsilon_i)}$, where $\epsilon_i$ are the
single-particle energies and $\delta$ is a Kronecker
delta constraint. For bosons $n_i = 0,1,\cdots,\infty$, and for
fermions $n_i = 0,1$. Writing the delta function as an integral
\begin{equation}
\label{2}
\delta (N - \sum_j n_j) = \frac{1}{2\pi} \int_0^{2\pi} d\varphi 
\exp (iN\varphi) \prod_i \exp(-in_i\varphi),
\end{equation}
one can easily perform the $n_i$ sums in Eq. (\ref{1}) to obtain
\begin{equation}
\label{3}
Z_N = \frac{1}{2\pi} \int_0^{2\pi} d\varphi \exp (iN\varphi) \left[
\prod_i \left[ 1 + \eta x_i \exp(-i \varphi) \right]^{\eta}
\right],
\end{equation}
where $\eta$ is $-1$ for bosons and $+1$ for fermions. Comparing
Eq. (\ref{3}) with the generating functions given in the Appendix A,
Eqs. (\ref{aa1}) and (\ref{aa2}), one immediately identifies the
generating functions of the symmetric functions inside the brackets in
the argument of the integral. For bosons one gets
\begin{equation}
\label{4}
Z_N = \frac{1}{2\pi} \int_0^{2\pi} d\varphi \exp (iN\varphi) \left[ 
\sum_M h_M(x) \exp(-i M \varphi) \right] = h_N (x),
\end{equation}
i.e. the complete symmetric function in the variables $x_i =
\exp{(-\beta \epsilon_i)}$. For fermions one gets
\begin{equation}
\label{4a}
Z_N = a_N(x), 
\end{equation}
i.e. the elementary symmetric function in the variables $x_i =
\exp{(-\beta \epsilon_i)}$. Hence the grand canonical partition
function is
\begin{equation}
\label{4b}
Z (\mu) = \sum_N h_N(x) \exp (- \beta \mu N) 
\end{equation}
for bosons and 
\begin{equation}
\label{4c}
Z (\mu) = \sum_N a_N(x) \exp (- \beta \mu N) 
\end{equation}
for fermions. (In this paper we will freely switch between the product
of the inverse temperature and the chemical
potential $\beta \mu$ and its analytic continuation $i \phi$). 

It is perhaps easier to understand the appearance of the symmetric
functions in the partition function by calculating  the quantity $Z_N$
using group characters. Group characters are the traces of the
representation matrices and can be expressed in terms of symmetric
functions \cite{weyl}. We are  interested in calculating the partition
function for a system of $N$  noninteracting particles with single
particle energies $\epsilon_i$ 
\begin{equation}
\label{5}
Z_N = {\rm Tr} [ \delta (N- \hat N) {\rm exp} ( - \beta \hat H)],  
\end{equation}
where the number operator is
\begin{equation}
\label{6}
\hat N= \sum_{i=1}^m c_i^{\dagger} c_i, 
\end{equation}
and the Hamiltonian is
\begin{equation}
\label{7}
\hat H = \sum_{i}^m  \epsilon_i c_i^{\dagger} c_i. 
\end{equation}
In these equations $c_i^{\dagger}$ and  $c_j$ are the creation and
annihilation operators for either bosons or fermions. 

Representations of continuous groups can be associated with Young
tableaux. Using $N$ bosons (fermions) distributed over m states, one
can construct completely symmetric (antisymmetric) irreducible
representations of the group U(m) associated with Young tableaux with
$N$ boxes in a row (column). Hence the delta  function in
Eq. (\ref{5}) restricts the trace to a given irreducible
representation. Since the quantity $\exp( - \beta \hat H)$ is a group
element,  the partition function in Eq. (\ref{5}) is simply the group
character in this representation. (Technically for  $\exp( - \beta
\hat H)$ to be a group element  one needs to perform a Wick rotation,
i.e. analytically continue the temperature to an imaginary variable).
In the theory of group representations, one can write the character of
any irreducible representation in terms of the eigenvalues of the
group element in  the fundamental representation, $x_i = \exp(-\beta
\epsilon_i)$. (In this representation the Hamiltonian is simply an $m
\times m$ matrix $H_{\rm fund}$ with  eigenvalues $\epsilon_i$). The
characters of the representation of U(m) associated with a single row
(column) Young tableaux are the complete (elementary) symmetric
functions of the eigenvalues of the group element in the fundamental
representation. It follows then that $Z_N$ is the complete symmetric
function in the variables $x_i = \exp{(-\beta \epsilon_i)}$ for bosons
and the elementary symmetric function in these variables for the
fermions. 

Using the generating functions of the symmetric functions,
Eqs. (\ref{aa1}) and (\ref{aa2}) of the Appendix A the grand canonical
partition functions in Eqs. (\ref{4b}) and (\ref{4c}) can be written
in the form 
\begin{equation}
\label{7a}
Z (\lambda) = \det \left[ 1 + \eta \lambda e^{- \beta \hat H}
\right]^{\eta}, 
\end{equation}
where $\lambda = e^{ - \mu}$ and $\eta$ is $-1$ for bosons and $+1$
for fermions as before. Using the relationship
\begin{equation}
\label{7b}
\det \hat A = \exp [ {\rm Tr} \log \hat A]
\end{equation}
which is valid for any operator $\hat A$ one can write Eq. (\ref{7a})
as 
\begin{eqnarray}
\label{7c}
Z (\lambda) &=& \sum_N Z_N \lambda^N \exp \left[ \eta {\rm Tr} \log
\left(  1 + \eta \lambda e^{- \beta \hat H} \right) \right] \nonumber 
\\ &=& \sum_k ( -1)^{k+1} \eta^k \frac{1}{k} \left( {\rm Tr} e^{ - k
\beta \hat H} \right) \lambda^k. 
\end{eqnarray}
Equating powers of $\lambda$ in both sides of the Eq. (\ref{7c}) one
can easily write down the recursion relation
\begin{equation}
\label{7d}
N Z_N = \sum_{k=1}^N k C_k Z_{N-k} ,
\end{equation}
where $C_k = ( -1)^{k+1} \eta^k \frac{1}{k} ( {\rm Tr} e^{ - k
\beta \hat H})$. Eq. (\ref{7d}) is the recursion function discussed in
Ref. \cite{borrmann1993}. In the studies of multiparticle
distributions the quantities $C_k$ are known as combinants
\cite{Kauffmann:1978vw,Balantekin:1991pf,Hegyi:1993ia}. 

If we have a mixed system of bosons and fermions we can write the
Hamiltonian to be
\begin{equation}
\label{8}
\hat H = \sum_{i=1}^k  \epsilon_i b_i^{\dagger} b_i +
\sum_{\alpha=1}^m  \epsilon_{\alpha} f_{\alpha}^{\dagger} f_{\alpha} 
\end{equation}
where the boson states are labeled by the Latin indices and the
fermionic states are labeled by the Greek indices. We take
$b_i^{\dagger}$ and  $b_i$ ($f_{\alpha}^{\dagger}$ and $f_{\alpha}$)
to be the creation and annihilation operators of the bosons (fermions)
and $\epsilon_i$ ($\epsilon_{\alpha}$) to be the single-particle
energies. Suppose we want to calculate the partition function $Z_N$
where the {\em total} number of bosons {\em and} fermions, $N=N_B+N_F$
is fixed. Using Eqs. (\ref{4}) and (\ref{4a}) one can easily write the
expression
\begin{equation}
\label{9}
Z_N = \sum_{n+\ell=N} h_n (x_B) a_{\ell} (x_F), 
\end{equation}
where $x_B$ represent the variables $\exp{(-\beta \epsilon_i)}$ of
bosons and $x_F$ represent the variables $\exp{(-\beta
\epsilon_{\alpha})}$ of fermions. The partition function in
Eq. (\ref{9}) was introduced in Ref. \cite{BahaBalantekin:1981qy}
where it was called the ``graded homogeneous symmetric function'' of
degree $N$ in the variables  $\exp{(-\beta \epsilon_i)}$ and
$\left[ - \exp{(-\beta \epsilon_{\alpha})} \right] $. It is the
character of the supergroup $U(k/m)$
\cite{BahaBalantekin:1981qy,BahaBalantekin:1981pp} 
(The Hamiltonian, Eq. (\ref{8}) is an element of the corresponding
superalgebra).      

\section{Partition Function for Particles with Internal Symmetries}

Whenever a quantum gas consists of non-interacting particles with an
internal symmetry then it is possible to write the grand canonical
partition function and then project onto a particular representation
of the Lie group associated with the symmetry in consideration. Even
though this approach was first introduced in
Ref. \cite{Redlich:1980bf} in the context of statistical bootstrap
models, it is in fact completely generally applicable
\cite{Turko:1981nr}. One writes the grand canonical partition
function, $Z$,
\begin{equation}
\label{10}
Z = {\rm Tr} \exp \left[ - \beta \left( \hat H - \sum \mu_i Q_i
\right) \right] ,
\end{equation}
where $Q_i$ are the conserved quantities of the system and $\mu_i$ are
the chemical potentials assigned to each of the relevant conservation
laws. In order to find the partition function $Z_r$ corresponding to a
given representation $r$ of the symmetry in consideration one simply
expands the grand canonical partition function in terms of the
characters $\chi_r$ of the associated Lie group:
\begin{equation}
\label{11}
Z (\mu_i)= \sum_r \frac{1}{d_r} \chi_r (\mu_i) Z_r .
\end{equation}
In Eq. (\ref{11}), $d_r$ is the dimension of the representation. Since
when the group variables are set to zero the character of a given
representation gives the dimension of this representation, $\chi_r
(\mu_i = 0 ) = d_r$, when all $\mu_i$ are set to zero Eq. (\ref{11})
gives the grand canonical partition function as a sum over all the
representations as it should. In fact Eqs (\ref{4b}) and (\ref{4c})
can be considered as special cases of Eq. (\ref{11}) since the
particle number is a conserved U(1) symmetry with characters $\exp ( -
i N \phi)$. Character expansions for various groups are readily
available in the literature
\cite{Balantekin:1984km,Balantekin:2000vn,Balantekin:2001id}. Such
expansions of 
the grand canonical partition function were utilized in a variety of
contexts from understanding the role of internal symmetries in $p
\overline p$ annihilation \cite{Muller:1982gd} to imposing color
neutrality in a quark-gluon plasma \cite{Elze:1986wv}.

In this section we show that symmetric functions are very useful  in
generating such expansions. To illustrate this let us introduce two
kinds of fermions which are the spin-up and spin down components of an
SU(2) algebra which can be the ordinary spin, or the isospin or a
pseudo-spin. For the purposes of fixing the notation we will call
these fermions protons (with creation and annihilation operators
$f^{\dagger}_{\alpha, +}, f_{\beta,+}$) and neutrons (with creation
and annihilation operators $f^{\dagger}_{\alpha, -}, f_{\beta,-}$) and
the symmetry isospin. We consider a dilute gas so that the
interactions between these particles can be ignored and assume that
they sit at the same energy-levels which may correspond to a mean
field: 
\begin{equation}
\label{12}
\hat H = \sum_{\alpha, \sigma = \pm} \epsilon_{\alpha}
f^{\dagger}_{\alpha, \sigma} f_{\alpha , \sigma} .
\end{equation}
It is easy to write down the generators of the corresponding SU(2)
algebra as
\begin{eqnarray}
\label{13a}
\hat T_3 &=& \frac{1}{2} \sum_{\alpha} \left[ f^{\dagger}_{\alpha, 
+ } f_{\alpha , + } - f^{\dagger}_{\alpha, - } f_{\alpha , -} \right] 
\nonumber \\ &=& \frac{1}{2} ( \hat N_+ - \hat N_- ) ,
\end{eqnarray}
and 
\begin{equation}
\label{13b}
\hat T_+ =  \sum_{\alpha} f^{\dagger}_{\alpha, + } f_{\alpha , -} =
(T_-)^{\dagger} .  
\end{equation}
In Eq. (\ref{13a}), $\hat N_+$ and $\hat N_-$ are the number of
protons and neutrons respectively. This algebra is easy to manipulate 
in the $| N_+, N_- \rangle$ basis. However, if we wish to find the
partition function corresponding to a particular value of isospin
$I$, we need to go to the $| I, m_I \rangle$ basis. 

Note that the third component of isospin $T_3$ is an additive,
conserved quantum number. Using Eqs. (\ref{1}), (\ref{3}), (\ref{4c})
we can write the grand canonical partition function for protons as
\begin{equation}
\label{14a}
\prod_{\alpha} \left( 1 + x_{\alpha} e^{i \phi / 2} \right) =
\sum_{N_+} a_{N_+} (x) \exp ( i N_+ \phi /2) , 
\end{equation} 
and for neutrons as
\begin{equation}
\label{14b}
\prod_{\alpha} \left( 1 + x_{\alpha} e^{ - i \phi / 2} \right) =
\sum_{N_-} a_{N_-} (x) \exp ( - i N_- \phi /2) . 
\end{equation} 
In writing Eqs. (\ref{14a}) and (\ref{14b}) we used the fact,
suggested by Eq . (\ref{13a}), that the
chemical potentials for protons and neutrons have opposite sign. Hence
the total grand canonical partition function of the system is 
\begin{equation}
\label{15}
Z = \prod_{\alpha} \left( 1 + x_{\alpha} e^{i \phi / 2} \right)
\left( 1 + x_{\alpha} e^{ - i \phi / 2} \right) = \sum_{N_+,N_-}
a_{N_+} a_{N_-} \exp [ i  (N_+ -N_-)\phi /2 ] . 
\end{equation} 
On the right side of the Eq. (\ref{15}) one recognizes the matrix
element of the group element $\exp (i T_3)$ in the $| N_+, N_-
\rangle$ basis. One can easily write down the  total grand canonical
partition function as 
\begin{equation}
\label{16}
Z = \prod_{\alpha} \left( 1 + x_{\alpha}^2 + x_{\alpha} {\rm Tr}\,\,
 U \right) ,
\end{equation} 
where $U$ is the group element $\exp (i \phi T_3)$ in the $I=1/2$
representation: 
\begin{equation}
\label{17}
U_{I=1/2} = \left(\begin{array}{cc} e^{i \phi/2} & 0 \\ 0 & e^{- i
\phi/2}  
\end{array} \right) . 
\end{equation}

We calculate the total grand canonical partition function of
Eq. (\ref{15}) using the character expansion formula of
Ref. \cite{Balantekin:2000vn} the results of which are summarized in
Appendix B. Letting $t_1 =  e^{i \phi / 2}$ and $t_2 =  e^{ - i \phi /
2}$ we define
\begin{equation}
\label{18}
G ( x, t ) \equiv \prod_{\alpha} \left( 1 + x_{\alpha} t \right) =
\sum_{N=0}^{\infty} a_N (x) t^N.
\end{equation} 
Using Eq. (\ref{ab5}) of Appendix B, we write the total grand
canonical partition function of Eq. (\ref{15}) as
\begin{equation}
\label{19}
\left( \prod_{i=1}^2 G(x,t_i) \right) = \sum_{n_1=0}^{\infty}
\sum_{n_2 =0}^{n_1} \det (a_{n_j +i -j}) \chi_{(n_1,n_2)} (U) .
\end{equation}
The equations in the Appendix B are given for $U(N)$. For the case of
$SU(2)$ we need to write characters of a given representation $I$. If
$n_2=0$, then $\chi_{(n_1,0)} = \chi_{I=n_1/2}$. However if $n_2 \neq
0$ then $\chi_{(n_1,n_2)} = \chi_{I=(n_1-n_2)/2}$.  Hence the desired
character expansion of the total grand canonical partition function 
is 
\begin{equation}
\label{20}
Z = \sum_{n=0}^{\infty} a_n \chi_{I=n/2} (U) + \sum_{n=0}^{\infty}
\sum_{m =1}^{n} (a_n a_m - a_{n+1} a_{m-1}) \chi_{I= (n-m)/2} (U) .
\end{equation}
Using Eq. (\ref{20}) one can for example write down the partition
function of the mixed system which corresponds to the total isospin
zero as 
\begin{equation}
\label{21}
Z_{I=0} = 1 + \sum_{n=1}^{\infty} (a_n^2 - a_{n+1} a_{n-1}) .
\end{equation}
In these equations $a_n$ is the elementary symmetric function of
degree $n$ in the variables $x_{\alpha} = \exp ( - \beta
\epsilon_{\alpha} )$. 

To illustrate the utility of Eq. (\ref{21}) let us consider the simple
case of a one-dimensional harmonic oscillator potential for which 
the $N$-particle partition functions can be
explicitly calculated (see for example Ref. \cite{schmidt2}). Setting
the zero-point energy of the harmonic oscillator to zero (i.e. $\hat H
= \hbar \omega \hat N$) we get
\begin{equation}
\label{21a}
a_N = x^{N(N-1)/2} \prod_{n=1}^N \frac{1}{(1-x^n)} \, ,
\end{equation}
where $x=\exp (- \beta \hbar \omega)$. 
Substituting Eq. (\ref{21a}) into Eq. (\ref{21}) we obtain
\begin{equation}
\label{21b}
Z_{I=0} = 1 + \sum_{n=1}^{\infty} a_n^2  \frac{1}{\sum_{m=0}^n x_m} ,
\end{equation}
where the factor $(\sum x_m)^{-1}$ projects the $I=0$ state out of a
state with $n$ proton-neutron pairs. 

In the example we just considered the individual particles transformed
like the fundamental representation of the internal symmetry group. We
next examine what happens if they transform like another
representation. Again to be specific we consider pions, which are
bosons that transform like the adjoint ($I=1$) representation of the
isospin group. We assume that the energies (either the free-particle
energies - i.e. we ignore the mass difference between the charged- and
the neutral-pion or the mean field energies) are again the
same. Introducing the creation operators $b^{\dagger}_{i,+},
b^{\dagger}_{i,0}$, and $b^{\dagger}_{i,-}$ for $\pi^+,\pi^0$, and
$\pi^-$ respectively the Hamiltonian is
\begin{equation}
\label{22}
\hat H = \sum_i \epsilon_i ( b^{\dagger}_{i,+} b_{i,+} +
b^{\dagger}_{i,0} b_{i,0} + b^{\dagger}_{i,-} b_{i,-} ), 
\end{equation}
and the generators of the appropriate $SU(2)$ algebra can be written
as 
\begin{equation}
\label{23}
\hat T_+  = \sum_i ( b^{\dagger}_{i,0} b_{i,-} + b^{\dagger}_{i,+}
b_{i,0} ) = (\hat T_-)^{\dagger} 
\end{equation}
and
\begin{equation}
\label{24}
\hat T_3  = \sum_i ( b^{\dagger}_{i,+} b_{i,+} - b^{\dagger}_{i,-}
b_{i,-}) .
\end{equation}

The total grand canonical partition function can again easily be
written as 
\begin{equation}
\label{25}
Z = \prod_i ( 1 - x_i e^{i \phi} )^{-1} ( 1 - x_i )^{-1} ( 1 - x_i
e^{- i \phi} )^{-1},
\end{equation}
where $x_i = \exp ( - \beta \epsilon_i)$. Note that the factor $1/2$
in the Eq. (\ref{13a}) is missing in Eq. (\ref{24}) since the additive
quantum number $T_3$ is $\pm 1/2$ for the nucleons, but $\pm 1$ or $0$
for pions. This also leads the lack of $1/2$ in the exponentials
multiplying the chemical potential in Eq. (\ref{25}). Indeed the total
grand canonical partition function for pions can be written as 
\begin{equation}
\label{26}
Z =  \det \left[ \prod_i \left[ 1 - x_i U_{I=1} \right]^{-1} \right], 
\end{equation}
where the determinant is taken in the isospin space and $U_{I=1}$ is
the matrix $\exp ( i \phi T_3)$ in the adjoint ($I=1$)
representation: 
\begin{equation}
\label{27}
U_{I=1} = \left(\begin{array}{ccc} e^{i \phi} & 0 & 0\\ 0 & 1 & 0 \\  
 0 & 0 & e^{- i \phi}  
\end{array} \right) . 
\end{equation}
To calculate the character expansion we proceed as before. Again using
the eigenvalues of the isospin in the {\em fundamental}
representation, $t_1 =  e^{i \phi / 2}$ and $t_2 =  e^{ - i \phi /
2}$, we define
\begin{equation}
\label{28}
G ( x, t ) \equiv \prod_i \left( 1 - x_i t^2 \right)^{-1} =
\sum_{N=0}^{\infty} A_N (x) t^N.
\end{equation} 
It follows from Eq. (\ref{aa1}) of Appendix A that $A_N = 0$ for odd
$N$, and $A_N = h_{N/2}$ for even $N$. Following the same steps as
before we write 
\begin{equation}
\label{29}
Z = \left[ \prod_i ( 1 - x_i )^{-1} \right] \sum_{n_1} \sum_{n_2}
\det (A_{n_j +i -j}) \chi_{(n_1,n_2)} (U) .
\end{equation} 
Going from $U(2)$ to $SU(2)$ and following similar steps as those
leading to Eq. (\ref{21}) we can write down the partition
function of the mixed system which corresponds to the total isospin
zero as 
\begin{equation}
\label{30}
Z_{I=0} = \left[ \prod_i ( 1 - x_i )^{-1} \right] \left[ 1 + 
\sum_{n=1}^{\infty} (h_n^2 -h_{n+1}) \right] .
\end{equation}

One may need to calculate the total partition function of a mixed
system of nucleons and pions for a particular value of isospin. To do
so we can proceed in a similar way. Starting with the total grand
canonical partition function:
\begin{equation}
\label{30a}
Z = \left[ \prod_{\alpha} (1 + x_{\alpha} t_1) (1 + x_{\alpha} t_2)
\right] \left[ \prod_i (1 - x_i t_1^2)^{-1} (1 - x_1)^{-1} (1 - x_i
t_2^2)^{-1} \right] .
\end{equation}
Defining the series
\begin{equation}
\label{30b}
\left[ \prod_{\alpha} (1 + x_{\alpha} t) \right] \left[ \prod_i (1 -
x_i t^2)^{-1} \right] = \sum_N B_N (x_{\alpha},x_i) t^N 
\end{equation}
we find that
\begin{equation}
\label{30c}
B_N (x_{\alpha},x_i)   = \sum_{N=n+m} a_n (x_{\alpha}) {\cal H}_m (
x_i) 
\end{equation}
where ${\cal H}_{2n} (x_i) = h_n (x_i)$ and ${\cal H}_{2n+1} (x_i) =
0, n=1,2,3, \cdots$. Using similar steps as those leading to
Eq. (\ref{29}) we get
\begin{equation}
\label{30d}
Z = \left[ \prod_i ( 1 - x_i )^{-1} \right] \sum_{n_1} \sum_{n_2}
\det (B_{n_j +i -j}) \chi_{(n_1,n_2)} (U) .
\end{equation} 

It is straightforward if not tedious to generalize the discussion in
this section to higher internal symmetries using the character
expansion formulae in Appendix B. For SU(N), $N \ge 2$, there are
$N-1$ mutually commuting operators (elements of the Cartan
subalgebra). These can be expressed in terms of number operators. Each
such operator is then associated with the analytic continuation of a
chemical potential. From the resulting group element one follows the
same procedure we just outlined. 

\section{Partition Functions for Interacting Systems}

In some cases it is possible to utilize the techniques discussed in
the previous sections to the investigation of some interacting
systems. Typically is the Hamiltonian $\hat H$ can be written as a sum
of the generators of an algebra, then $\exp ( - \beta \hat H)$ is an
element of the associated group and its trace (character) can be
calculated by powerful group-theoretical methods. To illustrate this
we will consider the Hamiltonian representing a two-level system, 
where each level has a $K$-fold degeneracy, given by 
\begin{equation}
\label{31}
\hat H = \frac{1}{2} \sum_{k=1}^K \left[ v \left( f^{\dagger}_{1,k}
f_{1,k} 
- f^{\dagger}_{1,k} f_{1,k} \right) + t\left( f^{\dagger}_{1,k}
f_{2,k} + f^{\dagger}_{2,k} f_{1,k} \right) \right] .   
\end{equation}
In Eq. (\ref{31}), $1$ and $2$ represent two different layers and 
the degeneracy of each level is indicated by the index $k$. An example
of such a system 
would be the single-particle Hamiltonian of a bilayer quantum Hall
system \cite{girvin}. In this case one works in a spherical geometry
and the $z$-projection of the orbital angular momentum of each
electron in the lowest Landau level, $k$, changes from $ - N_{\phi}/2$
to $N_{\phi}/2$, where $N_{\phi}$ is the number of flux quanta
penetrating the sphere.  The coefficients $v$ and $t$ are bias voltage
and the tunneling amplitude, respectively. 

To find the partition function we first diagonalize the Hamiltonian by
a Bogoliubov transformation \cite{bogol}: 
\begin{eqnarray}
\label{32}
F_{1,k} &=& \cos \theta f_{1,k} + \sin \theta f_{2,k}\,\, ,  \nonumber
\\  
F_{2,k} &=& - \sin \theta f_{1,k} + \cos \theta f_{2,k}\,\, .  
\end{eqnarray}
Note that $F_{i,k}, i=1,2$ such defined still satisfy the fermion
anticommutation relations. In addition, under the transformation in
Eq. (\ref{32}) the total number of particles is unchanged, i.e. 
\begin{equation}
\label{32a}
F^{\dagger}_{1,k} F_{1,k} + F^{\dagger}_{1,k} F_{1,k} =
f^{\dagger}_{1,k} f_{1,k} + f^{\dagger}_{1,k} f_{1,k} .
\end{equation}
By choosing 
\begin{eqnarray}
\label{33}
\cos \theta &=& \frac{v}{\sqrt{v^2+t^2}}, \nonumber \\
\sin \theta &=& \frac{t}{\sqrt{v^2+t^2}},
\end{eqnarray}
and
\begin{equation}
\label{34}
\epsilon = \sqrt{v^2+t^2},
\end{equation}
one can write down the Hamiltonian in Eq. (\ref{31}) in terms of the
quasi-fermion operators: 
\begin{equation}
\label{35}
\hat H = \frac{1}{2} \sum_{k=1}^K  \epsilon \left[ F^{\dagger}_{1,k} 
F_{1,k} - F^{\dagger}_{1,k} F_{1,k} \right].
\end{equation}

The total partition function can easily be computed as
$F^{\dagger}_{1,k}$ and $ F^{\dagger}_{2,k}$ create 
independent particles with energies $\epsilon /2$ and $- \epsilon /2$ 
respectively. The total possible number of both the ``upper-level''
and the ``lower-level'' particles are $K$. We get
\begin{equation}
\label{35a}
Z = \left( 1 + x_1 \right)^K \left( 1 + x_2 \right)^K, 
\end{equation}
where $x_1 \equiv \exp (- \beta \epsilon / 2 )$ and $x_2
\equiv \exp (+ \beta \epsilon / 2 )$. 
The partition function for a fixed number of particles can also be
similarly calculated. 
The partition function for $n$ ``upper-level'' particles
is given by applying Eq. (\ref{4a}) to $K$ degenerate levels with
energies $\epsilon /2$. (Note that in calculating the elementary
symmetric function the $k$ index in cannot repeat itself). Since 
\begin{equation}
\label{35b}
(1 + x_1  \lambda )^K = \sum_n \left(\begin{array}{c} K \\ n
\end{array} \right) x_1^n \lambda^n 
\end{equation}
with 
\begin{equation}
\label{35c}
\left(\begin{array}{c} K \\ n \end{array} \right) =
\frac{K!}{n!(K-n)!} 
\end{equation}
we get   
\begin{equation}
\label{36}
Z_n^+ = \left(\begin{array}{c} K \\ n  \end{array} \right) e^{- n
\beta \epsilon / 2 } .
\end{equation}
Similarly for the ``lower-level'' particles we get
\begin{equation}
\label{37}
Z_n^- = \left(\begin{array}{c} K \\ n  \end{array} \right) e^{+ n
\beta \epsilon / 2 } .
\end{equation}
Hence the $N$-particle partition function is given by
\begin{equation}
\label{38}
Z_N = \sum_{m+n=N} Z_n^+ \, Z_m^- \, . 
\end{equation}
To illustrate the dependence of the $N$-particle partition function on
the variables $x_1$ and $x_2$ we calculate $Z_N$ for the lowest
values of $N$. Even though $x_1$ and $x_2$ are inverses of each other
we will write their product explicitly to illustrate the underlying
structure. One gets 
\begin{equation}
\label{39}
Z_1 = K (x_1 +x_2), 
\end{equation}
\begin{equation}
\label{40}
Z_2 = \frac{K(K-1)}{2} (x_1^2 +x_2^2+x_1x_2) +  \frac{K(K+1)}{2}
(x_1x_2) , 
\end{equation}
\begin{eqnarray}
\label{41}
Z_3 &=& \frac{K(K-1)(K-2)}{3!} (x_1^3 + x_1^2 x_2 + x_1 x_2^2  +
x_2^3) \nonumber \\
&+&  \frac{K(K+1)(K-1)}{3} (x_1x_2) (x_1 +x_2) , 
\end{eqnarray}
\begin{eqnarray}
\label{42}
Z_4 &=& \frac{K(K-1)(K-2)(K-3)}{4!} (x_1^4 + x_1^3 x_2 + x_1^2 x_2^2 +
x_1 x_2^3 + x_2^4) \nonumber \\ &+&  \frac{K(K+1)(K-1)(K-2)}{8}
(x_1x_2) (x_1^2  + x_2^2+x_1x_2) + \frac{K^2(K^2-1)}{12} (x_1^2 x_2^2)
,   
\end{eqnarray}
and so on. In Eqs. (\ref{39}) through (\ref{42}) one notices the
appearance of both complete and elementary symmetric functions of $x_1$
and $x_2$. 

There is a much faster way to calculate these partition functions. 
Noting that the operators
\begin{equation}
\label{43}
\hat J_0 = \frac{1}{2} \sum_{k=1}^K  \left( f^{\dagger}_{1,k} 
f_{1,k} - f^{\dagger}_{1,k} f_{1,k} \right) ,
\end{equation}
and
\begin{equation}
\label{44}
\hat J_+ = f^{\dagger}_{1,k} f_{2,k} = (\hat J_-)^{\dagger} ,
\end{equation}
generate an SU(2) algebra one can write the Hamiltonian in
Eq. (\ref{31}) as an element of this algebra
\begin{equation}
\label{45}
\hat H = v \hat J_0 + \frac{t}{2} ( \hat J_+ + \hat J_- )
\end{equation}
As a result $\exp ( - \beta \hat H )$ is an element of the
corresponding SU(2) group and the partition function 
is the sum of traces (characters) of all possible representations of
this group element. In the fundamental (two-dimensional or spinor)
representation this group element takes the form
\begin{equation}
\label{46}
e^{ - \beta \hat H} = \left(\begin{array}{cc} e^{ - \beta \epsilon /2}
 & 0 \\ 0 & e^{ +  \beta \epsilon /2}   
\end{array} \right) = \left(\begin{array}{cc} x_1 & 0 \\ 0 & x_2    
\end{array} \right) . 
\end{equation}
We want to express the total partition function of Eq. (\ref{35a}) in
terms of the characters of this SU(2) algebra. Noting
\begin{equation}
\label{47}
(1 + t )^K = \sum_n \left(\begin{array}{c} K \\ n \end{array} \right)
t^n  
\end{equation}
and Eq. (\ref{ab5}) of the Appendix B we can write  Eq. (\ref{35a}) 
as 
\begin{equation}
\label{48}
Z = \left( 1 + x_1 \right)^K \left( 1 + x_2 \right)^K =  \sum_{n_1=0}
\sum_{n_2 =0} \det \left[ \left(\begin{array}{c} K \\ n_j+i-j
\end{array} \right) 
 \right] \chi_{(n_1,n_2)} \left( e^{ - \beta \hat
H} \right) .
\end{equation}
Using $N$ particles one can construct those representations of the
SU(2) in Eqs. (\ref{43}) and (\ref{44}) where $N=n_1+n_2$. The easiest
way to see that is to consider the grand canonical partition function 
\begin{equation}
\label{48a}
Z = \left( 1 + \lambda x_1 \right)^K \left( 1 + \lambda x_2 \right)^K
=   \sum_{n_1=0} \sum_{n_2 =0} \det \left[ \left(\begin{array}{c} K \\
n_j+i-j \end{array} \right) \right] 
\det \left( h_{n_j+i-j} (\lambda x_1 ,  \lambda x_2 ) \right) .
\end{equation}
Since the complete symmetric function satisfy the condition
\begin{equation}
\label{48b}
h_n (\lambda x_1 ,  \lambda x_2 ) =  {\lambda}^n h_n ( x_1 , x_2 ) ,
\end{equation}
one can write the character in Eq. (\ref{48a}) as 
\begin{equation}
\label{48c}
\det \left( h_{n_j+i-j} (\lambda x_1 ,  \lambda x_2 ) \right) =
{\lambda}^{n_1 + n_2} \det \left( h_{n_j+i-j} ( x_1 , x_2 ) \right)
\end{equation}
and the proof follows. 

For $N=1$
using Eq. (\ref{ab1}) one has $\chi_{(1,0)} ( e^{ - \beta \hat H}) =
h_1(x_1,x_2) = x_1+x_2$ with the coefficient 
\begin{equation}
\label{49}
\left(\begin{array}{c} K \\ 1 \end{array} \right) = K ,
\end{equation}
i.e. the result given in Eq. (\ref{39}). For N=2, there are two
possibilities: $n_1=2,n_2=0$ and $n_1=1,n_2=1$. To the expansion in
Eq. (\ref{48}) these contribute the terms
\begin{equation}
\label{50}
\left(\begin{array}{c} K \\ 2 \end{array} \right) h_2 (x_1,x_2) +
\det \left(\begin{array}{cc} K & 1 \\ \frac{K(K-1)}{2} & K \end{array}
\right) [ h_1^2 (x_1,x_2) - h_2 (x_1,x_2) ], 
\end{equation}
which, after evaluating the determinants, gives Eq. (\ref{40}). One can
similarly calculate the coefficients of the $n_1 + n_2 = 3$ and $4$
terms in Eq. (\ref{48}) to obtain Eqs. (\ref{41}) and (\ref{42})
respectively. 

\section{Conclusions}

In this paper we showed that various partition functions for free
(either non-interacting or those that interact through one-body
Hamiltonians) particles can be written as a sum of one or more group
characters. This result is not surprising since the partition
function, being a trace, is invariant under the exchange of
single-particle energies, hence it can be written in terms of either
elementary or complete symmetric functions which form a complete basis
for any function that is symmetric under the exchange of its
variables. The resulting expressions are however very useful to
simplify the calculations of the partition functions for particles
that carry internal quantum numbers. 

One should emphasize that our techniques, being combinatorial in
nature, can be used to describe
particle multiplicity distributions even in those situations where one
does not start from a partition function or even when temperature is
not well-defined. Such applications range from pion multiplicity
distributions in heavy-ion collisions \cite{Kauffmann:1978vw} (where
a temperature can be defined for a system) to fermion-pair production
by a time-varying electric field 
\cite{Balantekin:1991aa,Seger:1996gs} (where the concept of
temperature is not introduced).

\section*{ACKNOWLEDGMENTS}

I thank Gernot Akemann and Bruce Barrett for bringing several
references to my attention. This work was supported in part by the
U.S. National 
Science Foundation Grant No.\ PHY-0070161  at the University of
Wisconsin, in part by the University of Wisconsin Research Committee
with funds granted by the Wisconsin Alumni Research Foundation, and in
part by the Alexander von Humboldt-Stiftung.  I am grateful to the
Max-Planck-Institut f\"ur Kernphysik and H.A. Weidenm\"uller for the
very kind hospitality. 

\appendix
\section*{A. Symmetric Functions}

The complete homogeneous symmetric function, $h_n (x)$, of degree $n$
in the arguments $x_i, i=1, \cdots, N$, is defined as the sum of the
products of the variables $x_i$, taking $n$ of them at a time. For
three variables $x_1,x_2,x_3$, the first few complete homogeneous
symmetric functions are
\[
h_1 (x) = x_1 + x_2 + x_3,
\]
\[
h_2(x) = x_1^2 + x_2^2 + x_3^2 + x_1 x_2 +  x_1 x_3 +  x_2 x_3,
\]
\[
h_3 (x) = \sum_i x_i^3 + \sum_{i \neq j} x_i^2 x_j + x_1x_2x_3.
\]
One can write the generating function for $h_n$ as
\begin{equation}
\label{aa1} 
\frac{1}{\prod_{i=1}^N (1 - x_i z)} = \sum_n h_n(x) z^n .
\end{equation}
If $x_1,x_2,x_3$ are the eigenvalues of a matrix $B$, the symmetric
functions can be written in terms of traces of powers of $B$, e. g. 
\[
h_1 (x) = Tr B, 
\]
\[
h_2 (x) = \frac{1}{2} \left[ Tr B^2 + (Tr B)^2 \right],
\]
and so on. 

The elementary symmetric functions, $a_n(x)$, are defined in a similar
way except that no $x_i$ can be repeated in any product. Again for
three variables $x_1,x_2,x_3$, the first few elementary symmetric
functions are 
\[
a_1 = h_1
\]
\[
a_2 = x_1 x_2 +  x_1 x_3 +  x_2 x_3,
\]
\[
a_3 =  x_1x_2x_3.
\]
One takes $a_n = 0$ if $n>N$ and $a_0 = h_0 = 1$. The generating
function for $a_n$ is given by 
\begin{equation}
\label{aa2} 
\prod_{i=1}^N (1 - x_i z) = \sum_n (-1)^n a_n(x) z^n .
\end{equation}
Note that, since the generating functions in Eqs. (\ref{aa1}) and
(\ref{aa2}) are inverses of each other one can write $h_k$ in terms of
$a_i, i=1,\cdots,k$ and vice versa. If one takes $x_i, i=1, \cdots, N$
to be eigenvalues of an $N \times N$ matrix $A$, then $a_N (x) = \det
A$ and $a_{N+1} (x) = 0 = a_{N+2} = \cdots$.

\section*{B. Character expansion formulae}

In this Appendix we summarize the character expansion formulae of
references \cite{Balantekin:1984km}, \cite{Balantekin:2000vn} and
\cite{Balantekin:2001id}.   
The representations of the $U(N)$ group are labeled by a partition
into $N$ parts: ($n_1,n_2,\cdots,n_N$) where $n_1 \ge n_2 \ge \cdots
\ge n_N$ (see for example Ref. \cite{weyl}). We denote the eigenvalues
of the group element $U$ in the fundamental representation by
$t_1,t_2, \cdots,t_N$. The character (trace of
the representation matrix) of the irreducible representation
corresponding to the partition ($n_1,n_2,\cdots,n_N$) of non-negative
integers is given by 
\begin{equation}
\label{ab1}
\chi_{(n_1,n_2,\cdots,n_N)} (U) = \det ( h_{n_j+i-j}) ,
\end{equation}
where $h_n$ is the complete symmetric function in the arguments $t_1,
\cdots, t_N$ of degree $n$. (For a review of its properties see
Appendix A). In these equations the arguments of the determinants
indicate the $(ij)$-th element of the matrix the determinant of which
is calculated.

To obtain the character expansion of Ref. \cite{Balantekin:2000vn}
consider the power series expansion 
\begin{equation}
\label{ab2}
G(x,t) = \sum_n A_n(x) t^n ,
\end{equation}
where $x$ stands for
all the parameters needed to specify the coefficients $A_n$. We assume
that this series is convergent for $|t|=1$. Given $N$ different $t$'s
wich we take eigenvalues of the matrix $U$: 
$t_1, \cdots, t_N$, we write down the character expansion (Eq. (2.17)
of Ref. \cite{Balantekin:2000vn}) using $N$ copies
of Eq.(\ref{ab2}) 
\begin{equation}
\label{ab3}
\left( \prod_{i=1}^N G(x,t_i) \right) = \sum_{m_1=0} \sum_{m_2 =0}
\cdots \sum_{m_{N-1}=0} \sum_{n_N} \det (A_{n_j +i -j}) \left(\det U
\right)^{n_N} \chi_{(\ell_1, \ell_2, \cdots, \ell_N)} (U),  
\end{equation}
where 
\begin{equation}
\label{ab4}
m_j = n_j - n_{j+1}, \> j = 1, \cdots, N-1. 
\end{equation}
If the sum over $n$ in the expression
Eq. (\ref{ab2}) we started with is restricted to the non-negative
values of $n$ (i.e., $A_n=0$ when $n<0$), then $n_N$ is non-negative
and we can absorb the term $(\det U)^{n_N}$ into the character to
obtain: 
\begin{equation}
\label{ab5}
\left( \prod_{i=1}^N G(x,t_i) \right) = \sum_{n_1=0} \sum_{n_2 =0}
\cdots \sum_{n_N=0} \det (A_{n_j +i -j}) \chi_{(n_1,n_2,\cdots,n_N)}
(U) .
\end{equation} 
Note that the summation in Eq. (\ref{ab5}) is over all irreducible
representations of $U(N)$, but in Eq. (\ref{ab4}) is restricted to
those representations where the number of boxes in the last row of the
Young Tableau is zero {\em and} an additional summation over $n_N$, 
which, in general can take both positive and negative values. For
further details see Ref. \cite{Balantekin:2000vn}.

\end{document}